\documentclass[twocolumn,preprintnumbers,prl]{revtex4}
\usepackage{amsfonts}
\usepackage[T1]{fontenc}
\usepackage{amsmath,amsbsy,amssymb,graphicx}
\usepackage{times}
\usepackage{amsmath}
\usepackage{amssymb}
\usepackage{graphicx}

\def\beginABC{\begin{subequations}}
\def\endABC{\end{subequations}}

\def\Section#1{\bigskip\noindent\textbf{\large#1}\par\noindent}
\let\section=\Section
\let\mathbf=\boldsymbol

\begin{document}

\title{{\Large Quantum Hall Effects with High Spin-Chern Numbers}\\
{\Large in Buckled Honeycomb Structure with Magnetic Order}}
\author{Motohiko Ezawa}
\affiliation{Department of Applied Physics, University of Tokyo, Hongo 7-3-1, 113-8656,
Japan }

\begin{abstract}
As a topological insulator, the quantum Hall (QH) effect is indexed by the
Chern and spin-Chern numbers $\mathcal{C}$ and $\mathcal{C}_{\text{spin}}$.
We have only $\mathcal{C}_{\text{spin}}=0$ or $\pm \frac{1}{2}$ in
conventional QH systems. We investigate QH effects in generic monolayer
honeycomb systems. We search for spin-resolved characteristic patterns by
exploring Hofstadter's butterfly diagrams in the lattice theory and fan
diagrams in the low-energy Dirac theory. The Chern and spin-Chern numbers
are calculated based on the bulk-edge correspondence in the lattice theory
and on the Kubo formula in the Dirac theory. It is shown that the spin-Chern
number can takes an arbitrary high value for certain QH systems in
coexistence with buckled structure and magnetic order. This is a new type of
topological insulators. Samples may be provided by silicene with
ferromagnetic order and transition-metal oxide with antiferromagnetic order.
\end{abstract}

\maketitle

%\date{}

The quantum Hall (QH) effect is one of the most fascinating phenomena in
condensed matter physics\cite{BookPrange,BookDasSarma,ZFE}. It is
characterized by the topological index\cite{TKNN}. Indeed, the integer QH
state at the filling factor $\nu =n$ has the Chern number $n$. The concept
of topological insulator stems from QH systems\cite{Hasan,Qi}. In general a
topological insulator without time reversal symmetry is indexed by the Chern
number $\mathcal{C}$ and the spin-Chern number $\mathcal{C}_{\text{spin}}$
when spin $s_{z}$ is a good quantum number\cite{Prodan}. We investigate the
spin-Chern number in QH systems. In the conventional monolayer QH system it
takes $1/2$ and $0$ alternately as up-spin and down-spin electrons fill
Landau levels successively. In the bilayer QH system it can be at most $1$
due to the layer degree of freedom\cite{ZFE}. There are no conventional QH
states with higher spin-Chern numbers.

The unconventional QH effect was discovered in graphene, which gives us a
new insight into the Dirac system\cite{Novoselov,Kim}. Dirac electrons are
ubiquitous in monolayer honeycomb systems. Recently, many honeycomb Dirac
materials with spin-orbit (SO) interactions have been discovered. Examples
are silicene\cite{GLayPRL,Kawai,Takamura}, perovskite transition-metal oxide
grown on [111] direction (TMO)\cite{Okamoto,Hu} and transition-metal
dichalcogenides (MX$_{2}$)\cite{Xiao,Feng,FengPNAS}. Silicene is a quantum
spin Hall insulator\cite{KaneMele,LiuPRL,LiuPRB} due to the SO interaction,
which is a particular type of a two-dimensional topological insulator.
Silicene is particularly interesting since we can control the Dirac mass
externally by applying electric field\cite{EzawaNJP}, exchange interactions%
\cite{EzawaQAHE,EzawaExM} and photo-irradiation\cite{EzawaPhoto}. The QH
effect in silicene has been studied based on the Dirac theory\cite{EzawaQHE}
as well as the lattice theory\cite{Beugeling}. We note that TMO and MX$_{2}$
are trivial insulators due to large staggered potentials or staggered
exchange interactions even though there are SO interactions.

Without external magnetic field, the spin-valley dependent Chern number
takes $\pm 1/2$. Thus we have only $16$ types of topological insulators\cite%
{EzawaExM}. In this classification, the Chern number can only be $%
-2,-1,0,1,2 $, while the spin-Chern number can be $-1,-1/2,0,1/2,1$.
However, once we switch on magnetic field, the Chern number can take every
integer values. In this work we investigate an intriguing possibility to
materialize QH states carrying higher spin-Chern numbers.

We have previously proposed a generic Hamiltonian for honeycomb systems\cite%
{EzawaExM}, which contains eight interaction terms mutually commutative in
the Dirac limit. Among them four contribute to the Dirac mass. The other
four contribute to the shift of the energy spectrum. We are able to make a
full control of the Dirac mass and the energy shift independently at each
spin and valley by varying these parameters, and materialize various
topological phases in silicene and other honeycomb systems. As a
particularly interesting system we can generate a state which contains only
down-spin electrons near the Fermi level at half filling. When they are
split into Landau levels in magnetic field they are expected to carry high
spin-Chern numbers.

Our main results are summarized as follows. We have explored Hofstadter's
butterfly diagrams\cite{Hatsugai93B,Hatsugai,Esaki,Sato,Hasegawa} to see a
global pattern of spin resolution in various honeycomb systems. We have also
explored fan diagrams in the Dirac theory to see a detailed pattern of spin
resolution in the low-magnetic field regime, where the magnetic field is of
the order of $10$ Tesla. We then calculate the Chern and spin-Chern numbers
based on the bulk-edge correspondence\cite{Hatsugai93B} in the lattice
theory and on the Kubo formula\cite{Gusynin95L} in the Dirac theory. They
show a perfect agreement in this regime. Our new finding is that there exist
indeed QH states which may have arbitrarily high spin-Chern numbers. They
would appear in silicene together with a proximity coupling to a
ferromagnet, and also in TMO in electric field. These materials are
characterized by the buckled honeycomb structure with ferromagnetic or
antiferromagnetic order.

\section{Main results}

The honeycomb lattice consists of two sublattices made of $A$ and $B$ sites.
We consider a buckled system with the layer separation $2\ell $ between
these two sublattices. The states near the Fermi energy are $\pi $ orbitals
residing near the $K$ and $K^{\prime }$ points at opposite corners of the
hexagonal Brillouin zone. The low-energy dynamics in the $K$ and $K^{\prime
} $ valleys is described by the Dirac theory. In what follows we use
notations $s_{z}=\uparrow \downarrow $, $t_{z}=A,B$, $\eta =K,K^{\prime } $
in indices while $s_{z}^{\alpha }=\pm 1$ for $\alpha =\uparrow \downarrow $, 
$t_{z}^{i}=\pm 1$ for $i=A$,$B$, and $\eta _{i}=\pm 1$ for $i=K,K^{\prime } $
in equations. We also use the Pauli matrices $\sigma _{a}$ and $\tau _{a}$
for the spin and the sublattice pseudospin, respectively.

We investigate the honeycomb system in perpendicular magnetic field $B$\ by
introducing the Peirls phase, $\Phi _{ij}=\frac{e}{h}\int_{\mathbf{r}_{i}}^{%
\mathbf{r}_{j}}\mathbf{A}\cdot d\mathbf{r}$, with $\mathbf{A}$ the magnetic
potential. Any hopping term from site $i$ to site $j$ picks up the phase
factor $e^{2\pi i\Phi _{ij}}$. The magnetic field is given by $B=2\Phi /3%
\sqrt{3}a^{2}$ in unit of $e/h$, where $a$ is the lattice constant and $\Phi 
$ is the magnetic flux penetrating one hexagonal area. Note that $\Phi =1$
implies $B=1.6\times 10^{5}$ Tesla in the case of graphene.

The relevant tight-binding model is given by\cite{EzawaExM}, 
\begin{align}
H& =-t\sum_{\left\langle i,j\right\rangle \alpha }e^{2\pi i\Phi
_{ij}}c_{i\alpha }^{\dagger }c_{j\alpha }-\mu \sum_{i\alpha }c_{i\alpha
}^{\dagger }c_{i\alpha }  \notag \\
& +i\frac{\lambda _{\text{SO}}}{3\sqrt{3}}\sum_{\left\langle \!\left\langle
i,j\right\rangle \!\right\rangle \alpha }s_{z}^{\alpha }e^{2\pi i\Phi
_{ij}}\nu _{ij}c_{i\alpha }^{\dagger }c_{j\alpha }+\lambda _{V}\sum_{i\alpha
}t_{z}^{i}c_{i\alpha }^{\dagger }c_{i\alpha }  \notag \\
& +\lambda _{\text{X}}\sum_{i\alpha }s_{z}^{\alpha }c_{i\alpha }^{\dagger
}c_{i\alpha }+\lambda _{\text{SX}}\sum_{i\alpha }t_{z}^{i}s_{z}^{\alpha
}c_{i\alpha }^{\dagger }c_{i\alpha },  \label{BasicHamil}
\end{align}%
where $c_{i\alpha }^{\dagger }$ creates an electron with spin polarization $%
\alpha $ at site $i$ in a honeycomb lattice, and $\left\langle
i,j\right\rangle /\left\langle \!\left\langle i,j\right\rangle
\!\right\rangle $ run over all the nearest/next-nearest-neighbor hopping
sites. The first term represents the nearest-neighbor hopping with the
transfer energy $t$. The second term represents the chemical potential. The
third term represents the SO coupling\cite{KaneMele} with $\lambda _{\text{SO%
}}$, where $\nu _{ij}=+1$ $(-1)$ if the next-nearest-neighboring hopping is
anticlockwise (clockwise) with respect to the positive $z$ axis. The fourth
term represents the staggered sublattice potential term\cite{KaneMele} with $%
\lambda _{V}$. It may be generated\cite{EzawaNJP} due to the buckled
structure by applying external electric field $E_{z}$, where $\lambda
_{V}=\ell E_{z}$. The fifth term is the mean exchange term\cite{EzawaQAHE}.
The sixth term represents the staggered exchange term\cite{EzawaExM} with
the difference $\lambda _{\text{SX}}$ between the $A$ and $B$ sites.

The low-energy Dirac Hamiltonian at the $K_{\eta }$ point is\cite{EzawaExM}%
\begin{align}
H_{\eta }=& v_{\text{F}}\left( \eta P_{x}\tau _{x}+P_{y}\tau _{y}\right) -\mu
\notag \\
& +\lambda _{\text{SO}}\eta \sigma _{z}\tau _{z}+\lambda _{V}\tau
_{z}+\lambda _{\text{X}}\sigma _{z}+\lambda _{\text{SX}}\sigma _{z}\tau _{z},
\label{TotalDirac}
\end{align}%
where $v_{\text{F}}=\frac{\sqrt{3}}{2}at$ is the Fermi velocity, and $%
P_{i}\equiv \hbar k_{i}+eA_{i}$ is the covariant momentum. We divide the
potential terms into two groups, one proportional to $\tau _{z}$ and the
other not. When the spin $s_{z}$ is diagonalized, they are given by $\Delta
_{s_{z}}^{\eta }\tau _{z}+\mu _{s_{z}}^{\eta }$, with 
\begin{equation}
\Delta _{s_{z}}^{\eta }=\eta s_{z}\lambda _{\text{SO}}+\lambda
_{V}+s_{z}\lambda _{\text{SX}}  \label{DiracMass}
\end{equation}%
and 
\begin{equation}
\mu _{s_{z}}^{\eta }=-\mu +s_{z}\lambda _{\text{X}}.  \label{DiracShift}
\end{equation}%
Here, $\Delta _{s_{z}}^{\eta }$ is the Dirac mass and $\mu _{s_{z}}^{\eta }$
shifts the energy spectrum.

Electrons make cyclotron motion under perpendicular magnetic field and fill
the energy levels. We evaluate the energy spectrum numerically based on the
tight-binding Hamiltonian (\ref{BasicHamil}) and analytically based on the
low-energy Dirac theory (\ref{TotalDirac}). To see a global pattern of spin
resolution we explore Hofstadter's butterfly diagrams\cite%
{Hatsugai93B,Hatsugai,Esaki,Sato,Hasegawa} in the lattice theory. To see a
detailed pattern of spin resolusion in the low-magnetic field regime we
explore fan diagrams in the Dirac theory. We then calculate the Chern and
spin-Chern numbers based on the bulk-edge correspondence\cite{Hatsugai93B}
in the lattice theory and on the Kubo formula\cite{Gusynin95L} in the Dirac
theory. They show a perfect agreement in the low-magnetic field regime. We
have applied these methods to examine QH states in various honeycomb
systems. Here we report two salient QH systems with high spin-Chern numbers.

\begin{figure}[t]
\centerline{\includegraphics[width=0.5\textwidth]{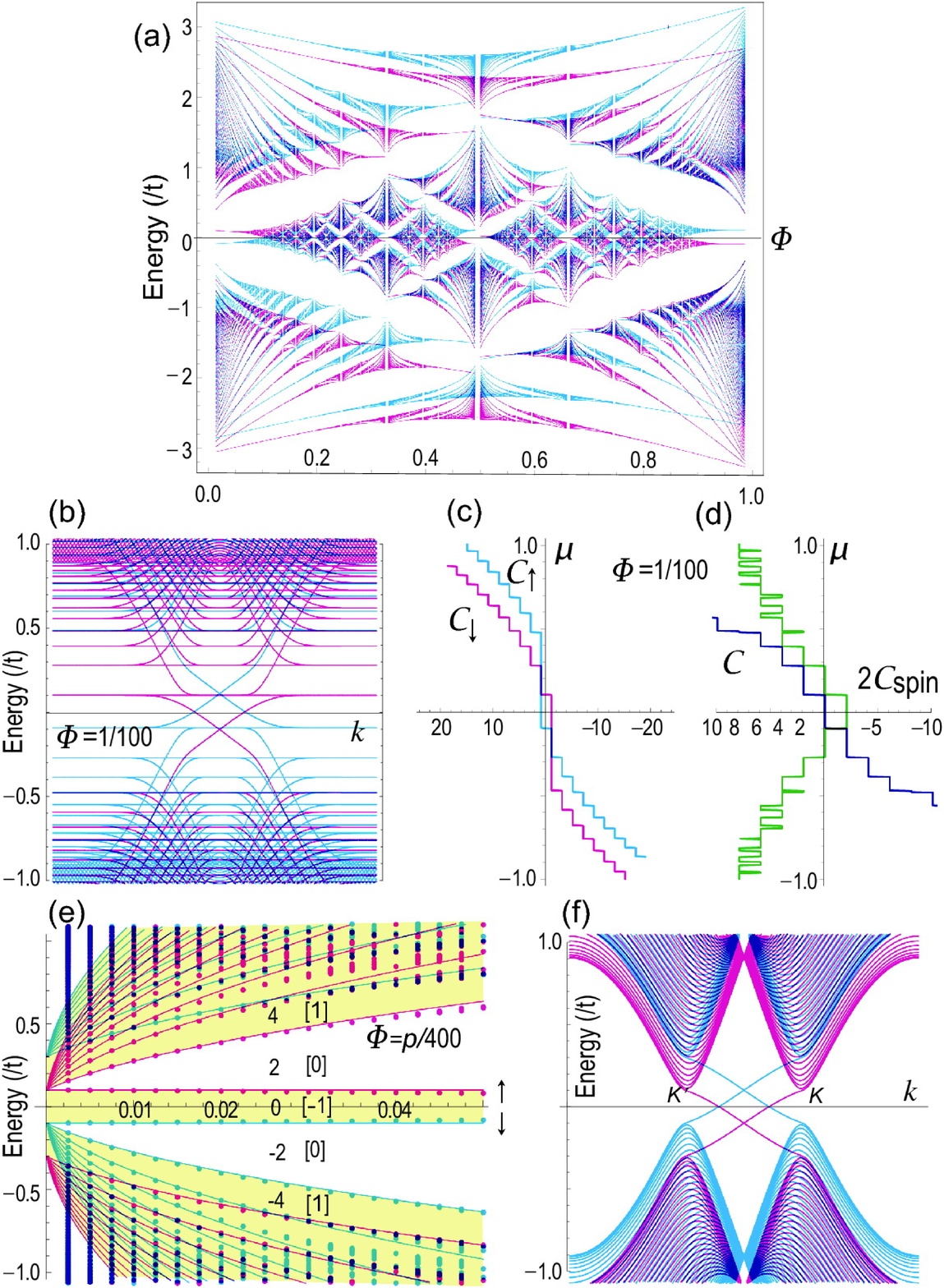}}
\caption{Silicene with ferromagnetic order. We have set $\protect\lambda _{%
\text{SO}}=0.2t$ and $\protect\lambda _{\text{X}}=0.1t$ for illustration.
The contribution from up(down)-spin electrons is shown in magenta (cyan). It
is indicated in violet when they are degenerate or almost degenerate. The
vertical axis is the energy in unit of $t$. (a) Spin-resolved Hofstadter's
diagram. The horizontal axis is the magnetic flux $\Phi $. We have taken $%
\Phi =p/q$ with $q\leq 100$. (b) The energy spectra of the honeycomb lattice
with zigzag edges for $\Phi =1/100$. The horizontal axis is the momentum $k$%
... (c) The topological number $\mathcal{C}_{s_{z}}$ calculated based on the
Dirac formula (\protect\ref{DiracCsz}) as a function of $\protect\mu $. The
horizontal axis is the topological number $\mathcal{C}_{s_{z}}$. (d) The
Chern and spin-Chern numbers $\mathcal{C}$ (blue) and $2\mathcal{C}_{\text{%
spin}}$ (green) derived based on (\protect\ref{ChernS}). (e) A closer look
of the Hofstadter's butterfly for $\Phi =p/400$ with $p=1,2,\cdots ,20$, and
the Landau levels (\protect\ref{SpectLL}) derived in the Dirac theory. The
horizontal axis is the magnetic flux $\Phi $. The yellow domain implies that
the spin-Chern number is nonzero. The Chern and spin-Chern numbers are
indicated as $\mathcal{C}$ $[\mathcal{C}_{\text{spin}}]$. (f) The band
structure of a zigzag nanoribbon at $\Phi =0$. Spins are polarized near the
Fermi level.}
\label{FigSiliZee}
\end{figure}

\textbf{Silicene with ferromagnetic order.}\textit{\ }The first example is
given by silicene with ferromagnetic order, where $t\approx 1.6$eV, $\lambda
_{\text{SO}}=3.9$meV. We introduce a ferromagnetic order by a proximity
coupling to a ferromagnet such as depositing Fe atoms to the silicene
surface or depositing silicene to a ferromagnetic insulating substrate\cite%
{Qiao,Tse,Yang}. Due to the ferromagnetic order ($\lambda _{\text{X}}\neq 0$%
), the energy levels of up-spin and down-spin electrons are shifted in
opposite directions, as illustrated in the band structure of a zigzag
nanoribbon [Fig.\ref{FigSiliZee}(f)]. Thus there appear only up-spin
electrons and down-spin holes near the Fermi level both for the $K$ and $%
K^{\prime }$ points at $\Phi =0$.

The Hofstadter diagram is displayed in Fig.\ref{FigSiliZee}(a). Fig.\ref%
{FigSiliZee}(e) is a closer look of the Hofstadter butterfly in the low
magnetic field regime ($\Phi <0.05$). It is well fitted by the spectrum (\ref%
{SpectLL}) obtained from the Dirac theory for $\Phi <0.01$. The fitting is
very good for the lowest and first Landau levels for a wide range of $\Phi $%
.. Two decomposed fans are visible due to the level shifting, as in the band
structure of a nanoribbon [Fig.\ref{FigSiliZee}(f)]. It is to be remarked
that all the energy levels are spin polarized and 2-fold degenerate with
respect to the valley degree of freedom.

Fig.\ref{FigSiliZee}(b) presents the edge-state analysis in magnetic field
at $\Phi =1/100$. According to the formula (\ref{BerryFormu}) we count the
number of edge modes, from which we derive the topological numbers $\mathcal{%
C}_{\uparrow }^{N}$ (magenta) and $\mathcal{C}_{\downarrow }^{N}$\ (cyan)
based on (\ref{ChernS}). On the other hand we may calculate them from the
Dirac formula (\ref{DiracCsz}), which we give in Fig.\ref{FigSiliZee}(c). We
can explicitly check that they agree one to another. The Chern and
spin-Chern numbers $\mathcal{C}^{N}$ (blue) and $\mathcal{C}_{\text{spin}%
}^{N}$\ (green) are given in Fig.\ref{FigSiliZee}(d).

It follows from Fig.\ref{FigSiliZee}(d) that the series of QH plateaux reads 
$\nu =0,\pm 2,\pm 4,\pm 6,\pm 8,\cdots $ at $\Phi =1/100$. On the other
hand, the spin-Chern number shows a complicated series since the up-spin and
down-spin fans cross as in Fig.\ref{FigSiliZee}(e). It changes by $\pm 1$
when a degenerate up(down)-spin level is crossed. We note that the QH state
at the Fermi energy has the topological indices $\mathcal{C}=0$ and $%
\mathcal{C}_{\text{spin}}=-1$ because silicene is a quantum spin-Hall state
without magnetic field.

We investigate an experimentally accessible regime, that is the low-magnetic
field limit $\Phi \lesssim 10^{-4}$ or $B\lesssim 16$ Tesla. The outstanding
feature is that only the up(down)-spin fan is present near the Fermi level
in the electron (hole) sector as in Fig.\ref{FigSiliZee}(e). The QH plateaux
reads $\nu =0,\pm 2,\pm 4,\pm 6,\pm 8,\cdots $ with 2-fold degeneracy in
every level. The spin-Chern number reads%
\begin{equation}
\mathcal{C}_{\text{spin}}=-1,0,1,2,3,\cdots .
\end{equation}%
The maximum value of $|\mathcal{C}_{\text{spin}}|$ increases as $\Phi $
becomes lower and $\lambda _{\text{X}}$ becomes larger. We may find QH
states with arbitrarily high spin-Chern numbers.

\begin{figure}[t]
\centerline{\includegraphics[width=0.5\textwidth]{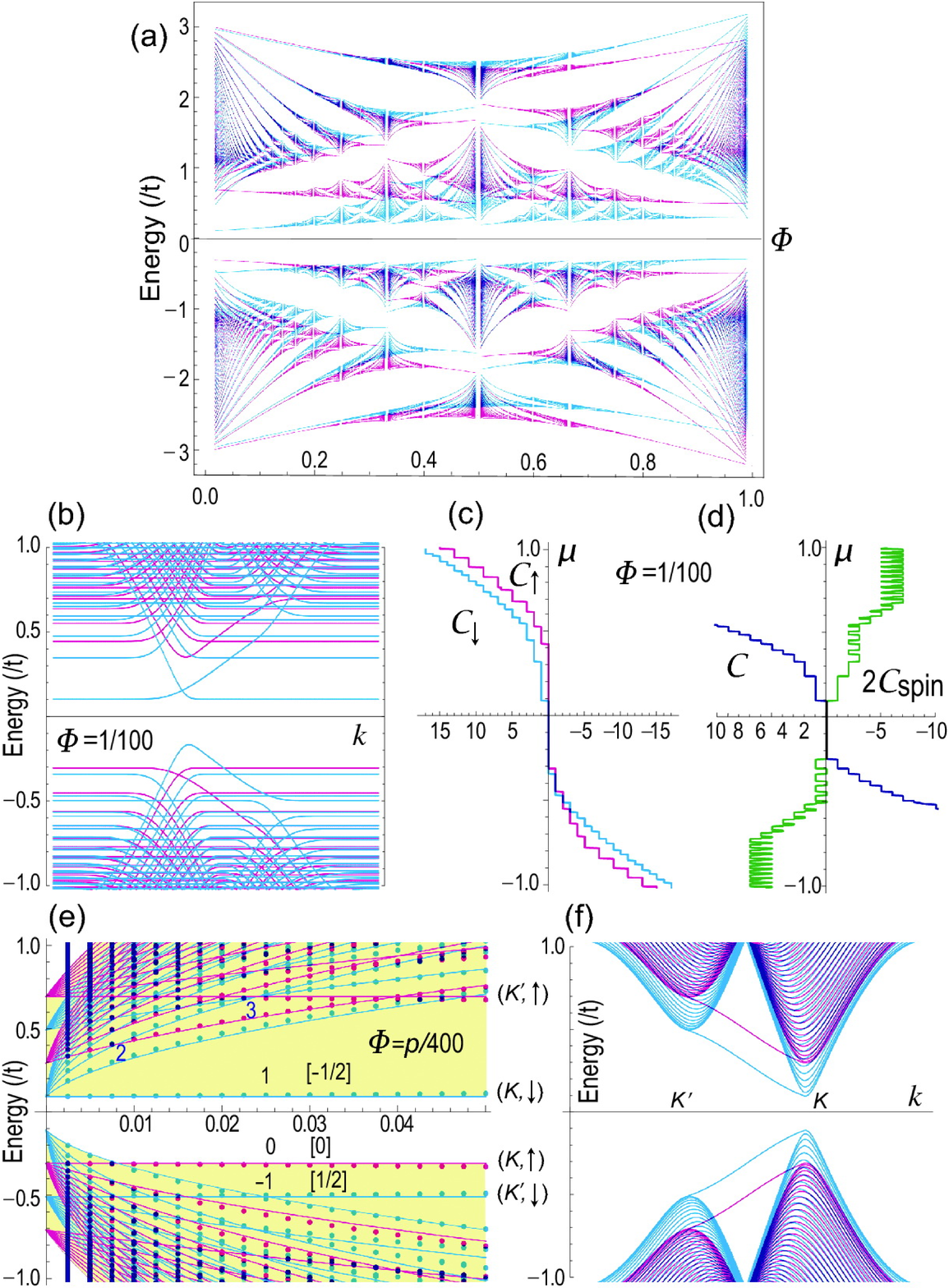}}
\caption{TMO in electric field. We have set $\protect\lambda _{\text{SO}%
}=0.2t$, $\protect\lambda _{V}=0.1t$ and $\protect\lambda _{\text{SX}}=0.4t$
for illustration. See the caption of Fig.\protect\ref{FigSiliZee}.}
\label{FigSQH}
\end{figure}

\textbf{Perovskite Transition-Metal Oxides.} The second example is given by
TMO, where $t\approx 0.2$eV, $\lambda _{\text{SO}}=7.3$meV, $\lambda
_{V}=\ell E_{z}$, $\lambda _{\text{SX}}=141$meV for LaCrAgO\cite{Hu}. A
salient property is that the material contains an intrinsic staggered
exchange effect $\varpropto \lambda _{\text{SX}}$. It has antiferromagnetic
order yielding Dirac mass. We can control the band structure by applying
electric field due to the buckled structure.\ When the electric field is off
($\lambda _{V}=0$), up-spin and down-spin electrons are degenerate. The
degeneracy is resolved as $\lambda _{V}$ increases [Fig.\ref{FigSQH}(f)],
and there appear only down-spin electrons and holes near the Fermi level
both for the $K$ and $K^{\prime }$ points at $\Phi =0$.

The Hofstadter diagram is displayed in Fig.\ref{FigSQH}(a). It is seen that
only down-spin electrons and holes exist dominantly near the Fermi level for
all values of $\Phi $. Fig.\ref{FigSQH}(e) is a closer look of the
Hofstadter butterfly in the low-magnetic field regime ($\Phi <0.05$). All
the four decomposed fans are visible due to four different masses, as is a
reflection of the band structure of a nanoribbon [Fig.\ref{FigSQH}(f)]. It
is to be remarked that all the energy levels are nondegenerate with respect
to the spin and valley degrees of freedom.

We present the edge-state analysis in Fig.\ref{FigSQH}(b), and the Chern and
spin-Chern numbers in Fig.\ref{FigSQH}(d). It follows from Fig.\ref{FigSQH}%
(d) that the series of QH plateaux reads $\nu =0,\pm 1,\pm 2,\pm 3,\pm
4,\cdots $ at $\Phi =1/100$. On the other hand, the spin-Chern number shows
a complicated series since the down-spin and up-spin fans cross as in Fig.%
\ref{FigSQH}(e). It changes by $\pm 1/2$ when a nondegenerate spin up (down)
level is crossed. Note that $\Phi =1/100$ implies $B\approx 140$ Tesla in
TMO.

We investigate the experimentally accessible regime, that is the low
magnetic field limit $\Phi \lesssim 10^{-3}$. The outstanding feature is
that only the down-spin fan is present near the Fermi level as in Fig.\ref%
{FigSQH}(e). The QH plateaux reads $\nu =0,\pm 1,\pm 2,\pm 3,\pm 4,\cdots $
with no degeneracy in every level. The spin-Chern number reads%
\begin{equation}
\mathcal{C}_{\text{spin}}=0,\mp \frac{1}{2},\mp 1,\mp \frac{3}{2},\mp
2,\cdots .
\end{equation}%
The maximum value of $|\mathcal{C}_{\text{spin}}|$ increases as $\Phi $
becomes lower and $E_{z}$ becomes larger. We may find QH states with
arbitrarily high spin-Chern numbers.

\section{Discussion}

We have shown that honeycomb systems allow QH systems with arbitrarily high
spin-Chern numbers. They give a new type of topological insulators. We have
presented two examples, silicene with ferromagnet order and TMO in electric
field. Similarly such QH states may occur in other honeycomb systems such as
boron-nitride, silicon carbide and transition metal dichalcogenides (MoS$%
_{2} $, etc.).

The condition for such QH states to appear is given essentially by the band
structure without magnetic field. We arrange the band structure to contain
only spin-polarized electrons of the same type near the Fermi level both at
the $K$ and $K^{\prime }$ points [Fig.\ref{FigSiliZee}(f) and Fig.\ref%
{FigSQH}(f)] by implementing appropriate magnetic order. For small magnetic
field, there are many Landau levels. When there are $N$ spin-polarized
energy levels, the maximum energy $\sqrt{2\hbar v_{\text{F}}^{2}eNB}$ must
be smaller than the energy gap $\Delta E$ between the two Dirac cones with
the opposite spins, where $\Delta E=|\Delta _{\uparrow }^{K}-\Delta
_{\downarrow }^{K}|$ or $\Delta E=|\Delta _{\uparrow }^{K^{\prime }}-\Delta
_{\downarrow }^{K^{\prime }}|$ with the Dirac mass (\ref{DiracMass}). The
maximum value of the spin-Chern number is given by $\left( \Delta E\right)
^{2}/4\hbar v_{\text{F}}^{2}eB$, when there exists no degeneracy in the
spectrum.

\section{Methods}

We have employed the following methods to make the analysis of QH systems in
various honeycomb systems and derive their Chern and spin-Chern numbers.

\textbf{Hofstadter Butterfly.} We compute the bulk band structure
numerically by applying periodic boundary conditions to the honeycomb
system. This requires that the magnetic flux $\Phi $ to be a rational
number, $\Phi =p/q$ ($p$ and $q$ are mutually prime integers). Then, the
system is periodic in both spatial directions. We use the Bloch theorem to
reduce the Schr\"{o}dinger equation to a $2q\times 2q$ matrix equation for
each $s_{z}=\uparrow \downarrow $, where the factor $2$ is due to the
sublattice ($A$,$B$) degrees of freedom. In so doing we choose a generalized
gauge of the one used in graphene\cite{Hatsugai93B} so as to include the
link connecting the next-nearest neighbor hopping sites. It is given in such
a way that the magnetic flux becomes $1/6$ for each isosceles triangle whose
two edges are given by the neighbor hopping. The resulting band structure is
the Hofstadters butterfly diagram. See Fig.\ref{FigSiliZee}(a) and Fig.\ref%
{FigSQH}(a).

\textbf{Fan Diagram.} We introduce a pair of ladder operators, 
\begin{equation}
\hat{a}=\frac{\ell _{B}(P_{x}+iP_{y})}{\sqrt{2}\hbar },\quad \hat{a}%
^{\dagger }=\frac{\ell _{B}(P_{x}-iP_{y})}{\sqrt{2}\hbar },  \label{G-OperaA}
\end{equation}%
satisfying $[\hat{a},\hat{a}^{\dag }]=1$, where $\ell _{B}=\sqrt{\hbar /eB}$
is the magnetic length. The Hamiltonian $H_{\eta }$ is block diagonal and
given by%
\begin{equation}
H_{\eta }=\left( 
\begin{array}{cc}
H_{\uparrow }^{\eta } & 0 \\ 
0 & H_{\downarrow }^{\eta }%
\end{array}%
\right) ,  \label{HamilBrockA}
\end{equation}%
with the diagonal elements being%
\begin{eqnarray}
H_{s_{z}}^{K} &=&\left( 
\begin{array}{cc}
\Delta _{s_{z}}^{K}+\mu _{s_{z}}^{K} & \hbar \omega _{\text{c}}\hat{a}%
^{\dagger } \\ 
\hbar \omega _{\text{c}}\hat{a} & -\Delta _{s_{z}}^{K}+\mu _{s_{z}}^{K}%
\end{array}%
\right) ,  \label{BlockEleme} \\
H_{s_{z}}^{K^{\prime }} &=&\left( 
\begin{array}{cc}
\Delta _{s_{z}}^{K^{\prime }}+\mu _{s_{z}}^{K^{\prime }} & -\hbar \omega _{%
\text{c}}\hat{a} \\ 
-\hbar \omega _{\text{c}}\hat{a}^{\dagger } & -\Delta _{s_{z}}^{K^{\prime
}}+\mu _{s_{z}}^{K^{\prime }}%
\end{array}%
\right)
\end{eqnarray}%
in the basis $\left\{ \psi _{A},\psi _{B}\right\} ^{t}$. Here, $\omega _{%
\text{c}}=\sqrt{2}v_{\text{F}}/\ell _{B}$ is the cyclotron frequency.

It is straightforward to solve the eigen equation of $H_{s_{z}}^{\eta }$.
The eigenvalues are\beginABC\label{SpectLL}%
\begin{equation}
E_{s_{z}}^{\eta }(N,\pm )=\mu _{s_{z}}^{\eta }\pm \sqrt{(\hbar \omega _{%
\text{c}})^{2}N+(\Delta _{s_{z}}^{\eta })^{2}},  \label{HighLL}
\end{equation}%
for $N=1,2,\cdots $, which depend on $\Phi $. We also have%
\begin{equation}
E_{s_{z}}^{\eta }(0)=\mu _{s_{z}}^{\eta }+\eta \Delta _{s_{z}}^{\eta },
\label{ZeroLL}
\end{equation}%
\endABC corresponding to $N=0$, which is independent of $\Phi $. The
eigenstate describes electrons when $E_{s_{z}}^{\eta }>0$ and holes when $%
E_{s_{z}}^{\eta }<0$. Note that, in the energy spectrum (\ref{HighLL}), $\pm 
$ corresponds to electrons or holes provided $\mu _{s_{z}}^{\eta }$ is zero
or sufficiently small.

We refer to each energy spectrum $E_{s_{z}}^{\eta }$ as a fan. There are
four fans indexed by valley $K_{\eta }$ and spin $s_{z}$. Each fan consists
of two parts, one for electrons and the other for holes. These two parts are
connected at one pivot when $\Delta _{s_{z}}^{\eta }=0$, and otherwise one
fan has two pivots. The separation between these two pivots is given by $%
2\Delta _{s_{z}}^{\eta }$, while the average distance of the two pivots from
the Fermi level is given by $\mu _{s_{z}}^{\eta }$. Let us call the energy
level (\ref{ZeroLL}) the lowest Landau level. In this convention there
exists one lowest Landau level in each fan. Thus there are four lowest
Landau levels in one fan diagram. It presents a picturesque illustration of
the Chern and spin-Chern numbers: See Fig.\ref{FigSiliZee}(e) and Fig.\ref%
{FigSQH}(e).

\textbf{Topological charges and conductance.} We consider the QH system at
the filling $\nu =N$. As a topological insulator it is indexed by a set of
two topological charges, the Chern number $\mathcal{C}^{N}$ and the
spin-Chern number $\mathcal{C}_{\text{spin}}^{N}$ given by\cite{Hasan,Qi}%
\begin{equation}
\mathcal{C}^{N}=\mathcal{C}_{\uparrow }^{N}+\mathcal{C}_{\downarrow
}^{N},\qquad \mathcal{C}_{\text{spin}}^{N}=\frac{1}{2}(\mathcal{C}_{\uparrow
}^{N}-\mathcal{C}_{\downarrow }^{N}),  \label{ChernS}
\end{equation}%
where $\mathcal{C}_{s_{z}}^{N}$ is the summation of the Berry curvature in
the momentum space over all occupied states of electrons at $\nu =N$ with $%
s_{z}=\uparrow \downarrow $. The charge-Hall and spin-Hall conductivities
are given by using the TKNN formula\cite{TKNN}, $\sigma _{xy}=(e^{2}/2\pi
\hbar )\mathcal{C}^{N}$, and $\sigma _{xy}^{\text{spin}}=(e/2\pi \hbar )%
\mathcal{C}_{\text{spin}}^{N}$.

\textbf{Buld-edge correspondence}\textit{.} The most convenient way to
determine the topological charge in the lattice formulation is to employ the
bulk-edge correspondence\cite{Hatsugai93B}. The edge-state analysis can be
performed for a system with boundaries such as a cylinder. When solving the
Harper equation on a cylinder, the spectrum consists of bulk bands and
topological edge states. See Fig.\ref{FigSiliZee}(b) and Fig.\ref{FigSQH}%
(b). We typically find a few edge states within the bulk gaps, some of which
cross the gap from one bulk band to another. Each edge state contributes one
unit to the quantum number $\mathcal{C}_{s_{z}}^{N}$ for each $%
s_{z}=\uparrow \downarrow $ at the filling $\nu =N$. More precisely, in
order to evaluate $\mathcal{C}_{s_{z}}^{N}$, we count the edge states,
taking into account their location (right or left edges) and direction (up
or down) of propagation\cite{Hatsugai93B}. The location of each state is
derived by computing the wave function, while the direction of propagation
can be obtained from the sign of its momentum derivative $dE/dk$, with $k$
the momentum parallel to the edge. We focus on one edge. Edge states with
opposite directions contribute with opposite signs. The resultant formula
reads%
\begin{equation}
\mathcal{C}_{s_{z}}^{N}=N_{\text{u}}^{s_{z}}-N_{\text{d}}^{s_{z}},
\label{BerryFormu}
\end{equation}%
where $N_{\text{u}}^{s_{z}}$ and $N_{\text{d}}^{s_{z}}$ denote the number of
up- and down-moving states with spin $s_{z}$, respectively, at the right
edge.

\textbf{Kubo formula.} We use the Kubo formulation in the Dirac theory to
derive the Hall conductivity for each spin $s_{z}$ in each valley $K_{\eta }$%
.. Such a formula has been derived for graphene\cite{Gusynin95L}. We may
generalize it to be applicable to the Dirac system (\ref{TotalDirac}),%
\begin{align}
\mathcal{C}_{s_{z}}(\mu )=& \frac{1}{4}\sum_{\eta }\Big[\tanh \frac{\mu
_{s_{z}}^{\eta }+\Delta _{s_{z}}^{\eta }}{2k_{\text{B}}T}+\tanh \frac{\mu
_{s_{z}}^{\eta }-\Delta _{s_{z}}^{\eta }}{2k_{\text{B}}T}\Big]  \notag \\
& +\frac{1}{2}\sum_{\eta ,n=1}^{\infty }\Big[\tanh \frac{E_{s_{z}}^{\eta
}(n,+)}{2k_{\text{B}}T}+\tanh \frac{E_{s_{z}}^{\eta }(n,-)}{2k_{\text{B}}T}%
\Big]  \label{DiracCsz}
\end{align}%
for each spin $s_{z}$. It is straightforward to calculate this as a function
of the chemical potential $\mu $ with the use of formulas (\ref{DiracMass})
and (\ref{DiracShift}) we obtain curves $\mathcal{C}_{\uparrow }(\mu )$ and $%
\mathcal{C}_{\downarrow }(\mu )$ in Fig.\ref{FigSiliZee}(c) and Fig.\ref%
{FigSQH}(c).

\section{Acknowledgements}

I am very much grateful to N. Nagaosa for many fruitful discussions on the
subject. This work was supported in part by Grants-in-Aid for Scientific
Research from the Ministry of Education, Science, Sports and Culture No.
22740196.

\section{Additional information}

\textbf{Competing financial interests:} The author declares no competing
financial interests.


\begin{thebibliography}{99}
\bibitem{BookPrange} R.E. Prange and S.M. Girvin (eds), \textit{The Quantum
Hall Effect} (Springer, 1990) 2nd edition.

\bibitem{BookDasSarma} S. Das Sarma and A. Pinczuk (eds), \textit{%
Perspectives in Quantum Hall Effects} (Wiley, 1997).

\bibitem{ZFE} Z. F. Ezawa, \textit{Quantum Hall Effects: Recent Theoretical
and Experimental Developments} (World Scientific, 2013) 3rd edition.

\bibitem{TKNN} D. J. Thouless, M. Kohmoto, M. P. Nightingale, and M. den
Nijs, Quantized Hall conductance in a two-dimensional periodic potential.
Phys. Rev. Lett. \textbf{49}, 405 (1982).

\bibitem{Hasan} M.Z Hasan and C. Kane, Colloquium: Topological insulators.
Rev. Mod. Phys. \textbf{82}, 3045 (2010).

\bibitem{Qi} X.-L. Qi and S.-C. Zhang, Topological insulators and
superconductors. Rev. Mod. Phys. \textbf{83}, 1057 (2011).

\bibitem{Prodan} E. Prodan, Robustness of the spin-Chern number. Phys. Rev.
B 80, 125327 (2009).

\bibitem{Novoselov} K. S. Novoselov, A. K. Geim, S. V. Morozov, D. Jiang, M.
I. Katsnelson, I. V. Grigorieva, S. V. Dubonos and A. A. Firsov,
Two-dimensional gas of massless Dirac fermions in graphene. Nature 438, 197
(2005).

\bibitem{Kim} Y. Zhang, Y. W. Tan, H. L. Stormer and P. Kim, Experimental
observation of the quantum Hall effect and Berry's phase in graphene. Nature
438, 201 (2005).

\bibitem{GLayPRL} P. Vogt, , P. De Padova, C. Quaresima, J. A., E.
Frantzeskakis, M. C. Asensio, A. Resta, B. Ealet and G. L. Lay, Silicene:
Compelling experimental evidence for graphenelike two-dimensional silicon.
Phys. Rev. Lett. \textbf{108}, 155501 (2012).

\bibitem{Kawai} C.-L. Lin, R. Arafune, K. Kawahara, N. Tsukahara, E.
Minamitani, Y. Kim, N. Takagi, M. Kawai, Structure of silicene grown on
Ag(111). Appl. Phys. Express 5, Art No. 045802 (2012) .

\bibitem{Takamura} A. Fleurence, R. Friedlein, T. Ozaki, H. Kawai, Y. Wang,
and Y. Yamada-Takamura, Experimental evidence for epitaxial silicene on
diboride thin films. Phys. Rev. Lett. \textbf{108}, 245501 (2012).

\bibitem{Okamoto} D. Xiao, W. Zhu, Y. Ran, N. Nagaosa and S. Okamato,
Interface engineering of quantum Hall effects in digital transition metal
oxide heterostructures. Nature Comm. \textbf{2}, 596 (2011).

\bibitem{Hu} Q.-F. Liang, L.-H. Wu, X. Hu, Electrically Tunable Topological
State in [111] Perovskite materials with antiferromagnetic exchange field.
cond-mat/arXiv:1301.4113

\bibitem{Xiao} D. Xiao, G.-B. Liu, W. Feng, X. Xu, and W. Yao, Coupled spin
and valley physics in monolayers of MoS2 and other group-VI dichalcogenides.
Phys. Rev. Lett. \textbf{108}, 196802 (2012).

\bibitem{Feng} T. Cao, G. Wang, W. Han, H. Ye, C. Zhu, J. Shi, Q. Niu, P.
Tan, E. Wang, B. Liu and J. Feng, Valley-selective circular dichroism of monolayer molybdenum disulphide, Nature Communications \textbf{3}, 887
(2012).

\bibitem{FengPNAS} X. Li, T. Cao, Q. Niu, J. Shin and J. Feng, Coupling the
valley degree of freedom to antiferromagnetic order. PNAS \textbf{110}, 3738
(2013)

\bibitem{KaneMele} C. L. Kane and E. J. Mele, Quantum spin Hall effect in
graphene. Phys. Rev. Lett. \textbf{95}, 226801 (2005).

\bibitem{LiuPRL} C.-C. Liu, W. Feng, and Y. Yao, Quantum spin Hall effect in
silicene and two-dimensional germanium. Phys. Rev. Lett. \textbf{107},
076802 (2011).

\bibitem{LiuPRB} C.-C. Liu, H. Jiang, and Y. Yao, Low-energy effective
Hamiltonian involving spin-orbit coupling in silicene and two-dimensional
germanium and tin. Phys. Rev. B, \textbf{84}, 195430 (2011).

\bibitem{EzawaNJP} M. Ezawa, Topological insulator and helical zero mode in
silicene under inhomogeneous electric field. New J. Phys. 14, 033003 (2012).

\bibitem{EzawaQAHE} M. Ezawa, Valley-polarized metals and quantum anomalous
Hall effect in silicene. Phys. Rev. Lett \textbf{109}, 055502 (2012).

\bibitem{EzawaExM} M. Ezawa, Spin-valleytronics in silicene:
quantum-spin-quantum-anomalous Hall insulators and single-valley semimetals.
Phys. Rev. B \textbf{87}, 155415 (2013)

\bibitem{EzawaPhoto} M. Ezawa, Photo-induced topological phase transition
and single Dirac-cone state in silicene. Phys. Rev. Lett. 110, 026603 (2013).

\bibitem{EzawaQHE} M. Ezawa, Quantum Hall effects in silicene. J. Phys. Soc.
of Jpn 81, 064705 (2012).

\bibitem{Beugeling} W. Beugeling, N. Goldman and C.M. Smith, Topological
phases in a two-dimensional lattice: Magnetic field versus spin-orbit
coupling. Phys. Rev. B \textbf{86}, 075118 (2012).

\bibitem{Hatsugai93B} Y. Hatsugai, Edge states in the integer quantum Hall
effect and the Riemann surface of the Bloch function. Phys. Rev. B \textbf{48%
}, 11851 (1993).

\bibitem{Hatsugai} Y. Hatsugai, T. Fukui and H. Aoki, Topological analysis
of the quantum Hall effect in graphene: Dirac-Fermi transition across van
Hove singularities and edge versus bulk quantum numbers. Phys. Rev. B 
\textbf{74} 205414 (2006).

\bibitem{Esaki} K. Esaki, M. Sato, M. Kohmoto, and B. I. Halperin, Zero
modes, energy gap, and edge states of anisotropic honeycomb lattice in a
magnetic field. Phys. Rev. B \textbf{80} 125405 (2009).

\bibitem{Sato} M. Sato, D. Tobe and M. Kohmoto, Hall conductance,
topological quantum phase transition, and the Diophantine equation on the
honeycomb lattice. Phys. Rev. B \textbf{78} 235322 (2008).

\bibitem{Hasegawa} Y. Hasegawa and M. Kohmoto, Quantum Hall effect and the
topological number in graphene. Phys. Rev. B \textbf{74} 155415 (2006).

\bibitem{Gusynin95L} V.P. Gusynin, S.G. Sharapov, Unconventional integer
quantum Hall effect in graphene. Phys. Rev. Lett. \textbf{95}, 146801
(2005): Transport of Dirac quasiparticles in graphene: Hall and optical
conductivities. Phys. Rev. B \textbf{73}, 245411 (2006).

\bibitem{Qiao} Z. Qiao, S. A. Yang, W. Feng, W.-K. Tse, J. Ding, Y. Yao, J.
Wang, and Q. Niu, Quantum anomalous Hall effect in graphene from Rashba and
exchange effects. Phys. Rev. B \textbf{82}, 161414 R (2010).

\bibitem{Tse} W.K. Tse Z. Qiao, Y. Yao, A. H. MacDonald, and Qian Niu,
Quantum anomalous Hall effect in single-layer and bilayer graphene. Phys.
Rev. B \textbf{83}, 155447 (2011).

\bibitem{Yang} Y. Yang, Z. Xu, L. Sheng, B. Wang, D.Y. Xing, and D. N.
Sheng, Time-reversal-symmetry-broken quantum spin Hall effect. Phys. Rev.
Lett. \textbf{107}, 066602 (2011).
\end{thebibliography}
\end{document}